# Design of PI Controller for Automatic Generation Control of Multi Area Interconnected Power System Using Bacterial Foraging Optimization

Naresh Kumari [#1], Nitin Malik [#2], A. N. Jha [*3], Gaddam Mallesham [#*4]
# Department of Electrical, Electronics and Communication Engineering,
The NorthCap University, Gurgaon, India
[1] nareshkumari@ncuindia.edu
[2] nitinmalik77@gmail.com
* Ex-Professor, Electrical Engineering, Indian Institute of Technology, New Delhi, India
[3] anjha@ee.iitd.ac.in
#* Department of Electrical Engineering, Osmania University, Hyderabad, India
[4] gm.eed.cs@gmail.com

*Abstract*— The system comprises of three interconnected power system networks based on thermal, wind and hydro power generation. The load variation in any one of the network results in frequency deviation in all the connected systems .The PI controllers have been connected separately with each system for the frequency control and the gains  (Kp  and Ki ) of all the controllers have been  optimized along with frequency bias (Bi) and speed regulation parameter ( Ri ) . The computationally intelligent techniques like bacterial foraging optimization (BFO) and particle swarm optimization (PSO) have been applied for the tuning of controller gains along with variable parameters Bi and Ri. The gradient descent (GD) based conventional method has also  been applied for optimizing the parameters Kp , Ki ,Bi and Ri .The frequency responses are obtained with all the methods . The performance index chosen is the integral square error (ISE). The settling time, peak overshoot and peak undershoot of all the frequency responses on applying three optimization techniques have been compared. It has been observed that the peak overshoot and peak undershoot significantly reduce with BFO technique followed by the PSO and GD techniques. While obtaining such optimum response the settling time is increased marginally with bacterial foraging technique due to large number of mathematical equations used for the computation in BFO. The comparison of frequency response using three techniques show the superiority of BFO over the PSO and GD techniques. The designing of the system and tuning of the parameters with three techniques has been done in MATLAB/SIMULINK environment.

**Keyword -** Bacterial foraging optimization; Particle swarm optimization; Gradient Descent ; Integral  square error; peak overshoot; peak undershoot; settling time; PI controller

## I. INTRODUCTION

One of the power system requirement is the stable operation in respect of voltage and frequency under varying load conditions. The variation in these parameters must lie within permissible range. The real power control help in achieving the frequency control and reactive power help in achieving the voltage control. The frequency deviation occurs in all the interconnected areas when they are connected through same tie line. The performance index has to be minimized continuously whenever there is frequency change .The performance index in the present case is chosen as the integral square error (ISE) which has to be minimized whenever there is frequency change using the computationally intelligent techniques. The load frequency controller is the proportional integral (PI) controller which is the most effective controller if it is properly tuned for proportional (Kp) and integral (Ki) gains and it is used in decentralized control mode as the interconnected areas are of unequal nature. The interconnected power system areas in most of the literature are of same nature but in actual scenario this trend is changing due to the increasing scarcity of non renewable energy generation sources. The interconnected systems are of different nature and unequal ratings which are connected with a tie line. The proper tuning of  regulation parameter (Ri) and frequency bias (Bi)  of these power systems is also a tedious task under varying load conditions [1].

The three tie line connected systems in the present network comprises of thermal, wind and hydro power generation systems. The gains of three controllers and variable parameters of the systems are optimized by computationally intelligent techniques like BFO, PSO and as well as classical gradient descent (GD) method [2] .The effectiveness of these methods depends upon the convergence, precision and robustness of the technique. There are chances that the solution parameter gets trapped in local minima in case of GD method and it does not provide the optimum solution. PSO is an evolutionary search algorithm which depends upon the concept of bird flocking and as it is an algorithm based on multi-dimensional search space so the chances of local minimum trapping is very less .Still there are some deficiencies in PSO which are overcome in more advanced bacterial foraging based algorithm BFO [3]. The search space is wider in BFO and simultaneous optimization of several parameters is more accurate and give better results.

In the present work the brief overview of BFO technique is also given for better understanding of the method and the literature needed for PSO and GD algorithm is available in  ([4], [5]) . The problems of PSO which are related to premature convergence are eliminated in BFO[6].

## II. BACTERIAL FORAGING TECHNIQUE

BFO is an advanced search technique which comprises of the large number of operations in comparison to PSO for searching the optimum solution in search space. The BFO algorithm was developed by Passino [7] .In the BFO technique the foraging of E.Coli bacteria which is present in our intestine is the base for algorithm. The BFO method is divided into four sections as chemotaxis, swarming, reproduction and elimination and dispersal ([8], [9]).

A. *Chemotaxis*

The chemotaxis process is related to the movement of the bacteria based on the rotation of Flagella. If the movement is in predefined direction then it is called swimming otherwise tumbling. The process of tumbling is given as:

$$\theta^i(j+1,k,l) = \theta^i(j,k,l) + C(i)\phi(j) \tag{1}$$

Where $\theta^i(j,k,l)$ shows the ith bacterium having jth chemotactic level , kth reproductive step and lth elimination and dispersal step. C(i) is the step size of movement in random direction.

B. *Swarming*

In the swarming process the attraction signal is sent to all neighboring bacteria from the bacteria moving towards the best food position and thus the bacteria form the groups. These groups further move towards the bacteria group of high density. The swarming phenomenon can be represented as :

$$J_{cc}(\theta, P(j,k,l)) = \sum_{i=1}^{S_f} J_{cc}^i\left(\theta, \theta^i(j,k,l)\right)$$
$$= \sum_{i=1}^{S_f} \left[-d_{attract} \exp\left(-\omega_{attract} \sum_{m=1}^{p} (\theta_m - \theta_m^i)^2\right)\right] \tag{2}$$
$$+ \sum_{i=1}^{S_f} \left[h_{repelent} \exp\left(-\omega_{repelent} \sum_{m=1}^{p} (\theta_m - \theta_m^i)^2\right)\right]$$

The $J_{cc}(\theta, p(j,k,l))$ is cost function value which is to be added to the actual cost function (ISE) to minimize. 'Sf' represents the total number of bacteria and 'p' represents the total number of parameters to be optimized.

C. *Reproduction*

During this process the bacteria which are weak generally die and the healthy bacteria splits into two. This way the  bacteria population  will  remain constant.

D. *Elimination and Dispersal*

After some reproduction processes there can be some unknown event due to which some bacteria get killed and other are shifted to new locations. These new food positions may or may not favour the chemotaxis process. The mathematical advantage of this process is that the premature solution and local optima is avoided. These processes help to achieve the effective solution with BFO technique.

The flow chart of BFO technique is given in Fig. 1:

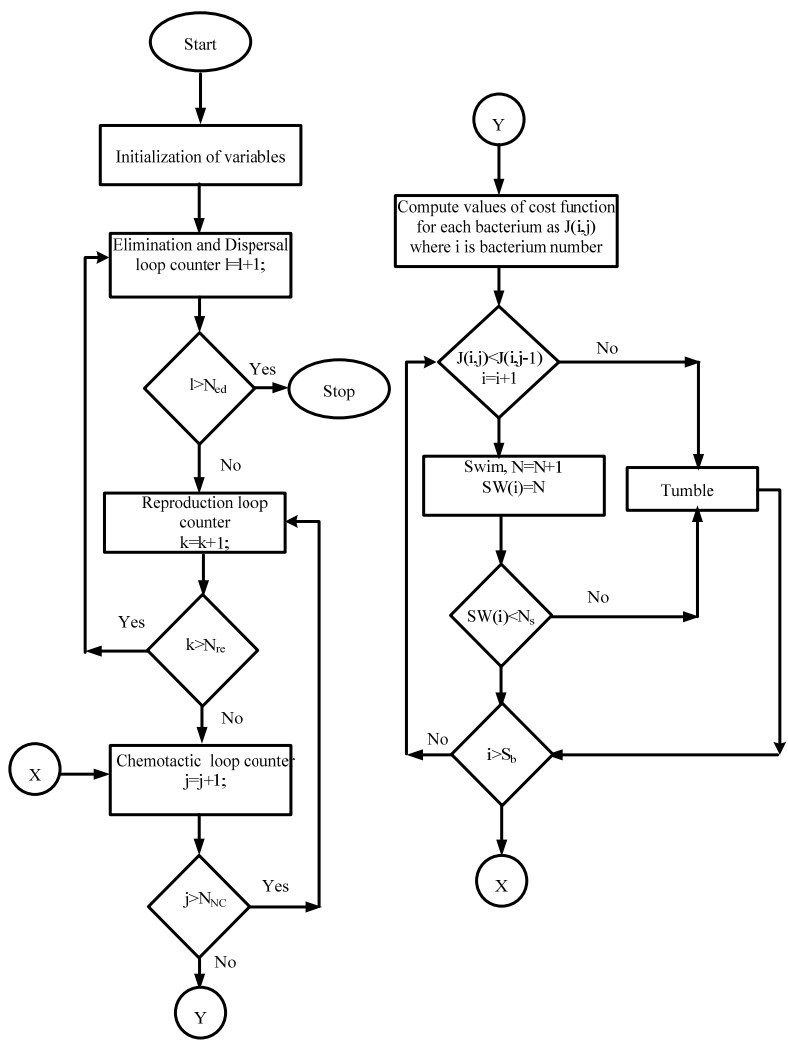

Fig. 1. Flow chart of the BFO algorithm.

The simulation work has been done considering the recent positions of variable parameters of system (Bi and Ri) and controller gains (Kp , Ki ) in search space as the bacteria .The various parameters are assigned with random values which lie between the upper and lower limits. The values of BFO parameters are taken as:

Search space dimension (p) =100

Chemotactic steps (Nc) =120

Total number of bacteria (s) =120

Limits the length of a swim (Ns) =30

Total number of steps of reproduction (Nre) =30

Total number of event for elimination and dispersal (Ned) =5

Number of bacteria reproductions per generation (Sr)=s/2

Probability each bacteria elimination (Ped) = 0.25

### III. MODELLING OF THE SYSTEM

The system comprises of thermal, wind and hydro power plants of ratings 2000 MW, 35 MW and 2000 MW respectively. These plant share the power through tie line ([10], [11]) .The reheat and generation rate constraint have been taken into account while designing the thermal system model ([12], [13], and [14]). The parameters of the three systems are given in Table I and II. The constant wind speed has been assumed for wind power plant. There are three separate PI controllers for each system. Transfer function models of all the three plants have been developed for the analysis of frequency response. For the simplicity of model and calculation work the models are made linear ([15],[16]). The interconnection of the three power systems using MATLAB/SIMULINK has been shown in Fig.2. The frequency bias and speed droop (Bi and Ri) control have been applied for wind plant . The damping ratio and natural frequency have been chosen carefully for the proper modelling of the wind plant. The active power of all plants must be controlled in such a manner that the stability is maintained.

The range for the values of Kp and Ki should be such that the stability of the system is maintained and the controllers are easily feasible to design with the optimized values. The performance index chosen is the integral square error (ISE) which can be given as :

$$ISE = \int \{(\Delta fi)2 + (\Delta Ptiei\text{-}j)2\}dt \qquad (3)$$

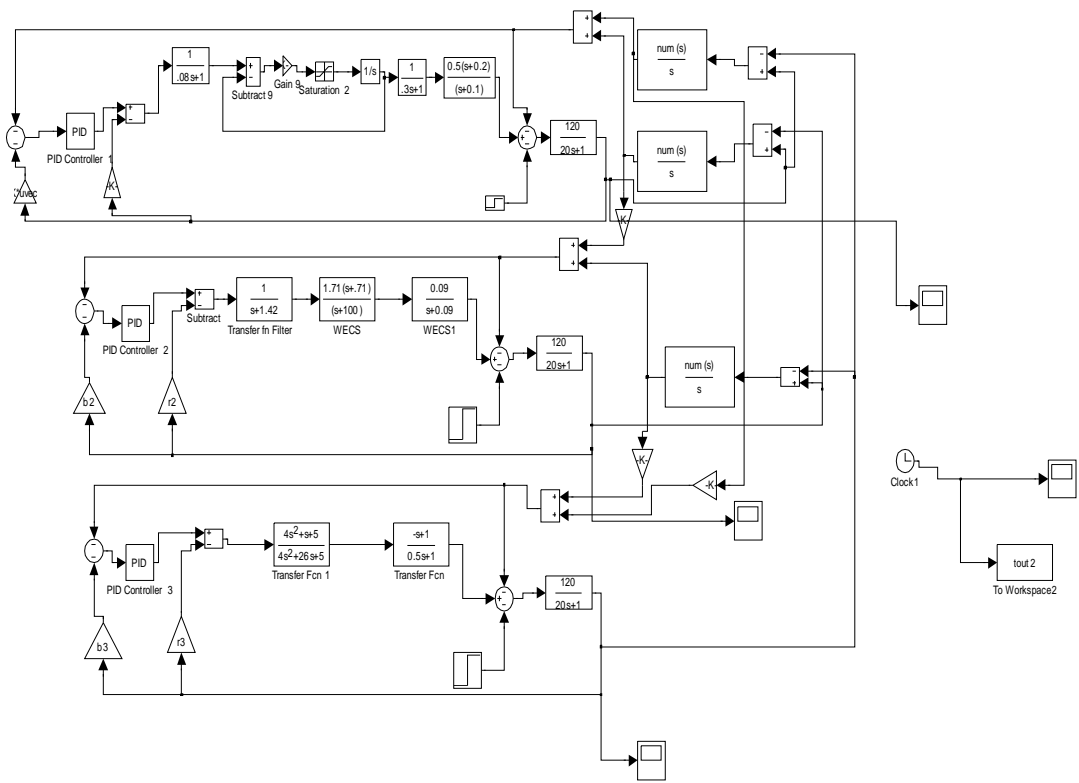

Fig 2.Three area interconnected reheat thermal, wind and hydropower system network

## IV. RESULTS AND DISCUSSION

The system under study has been subjected to a load change of 1 % at a time in all the three areas of the network and the frequency response has been observed in all the areas. Three different optimization techniques such as BFO, PSO and GD have been used for the tuning of PI controller gains Kp, Ki and variable parameters Bi and Ri of all the interconnected systems. In the BFO technique the limits of all the twelve variables for the three area system have been chosen and the variables of the algorithm are assigned the values as described in section II. The performance index has been subjected to BFO algorithm for minimization.

The PSO technique has been applied with the suitable variable values as initial values of PSO [17]. The GD technique has been applied by initializing the controller gains and system variables as zero. The BFO based PI controller gives the performance as shown in the different frequency response plots in Fig 3- Fig. 5. In the responses delf represents the frequency deviation, delf 11 is frequency response of power system area-1 for 1% load deviation in area-1, delf21 is frequency response of power system area-2 for 1% load deviation in area-1 and delf31 is frequency response of power system area-3 for 1% load deviation in area-1. When the same load variations are made in other areas the frequency changes but it is settled with properly tuned controller. The PSO and GD methods are also applied for the frequency response enhancement under the same load variation conditions and the frequency response is shown in Fig.6 - Fig. 8. The comparison of peak undershoot, peak overshoot and settling time for the frequency response has been done for all the techniques in Fig 9 – Fig 11.

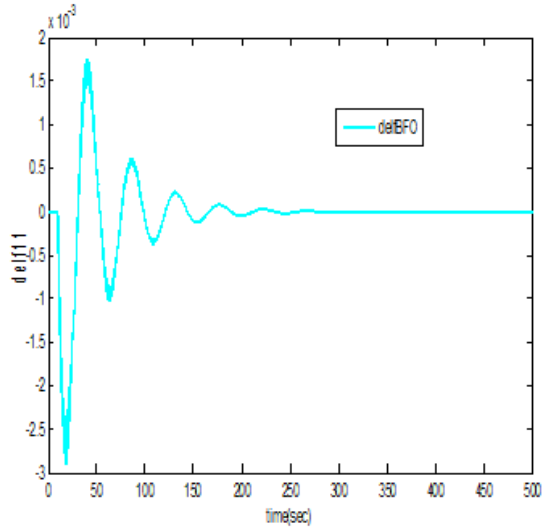

Fig 3. The frequency deviation in area 1 with 1% load change in area 1 with BFO technique

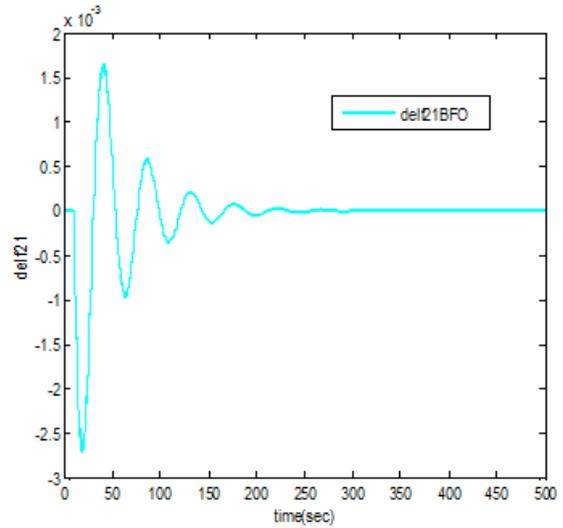

Fig 4. The frequency deviation in area 2 with 1% load change in area 1 with BFO technique

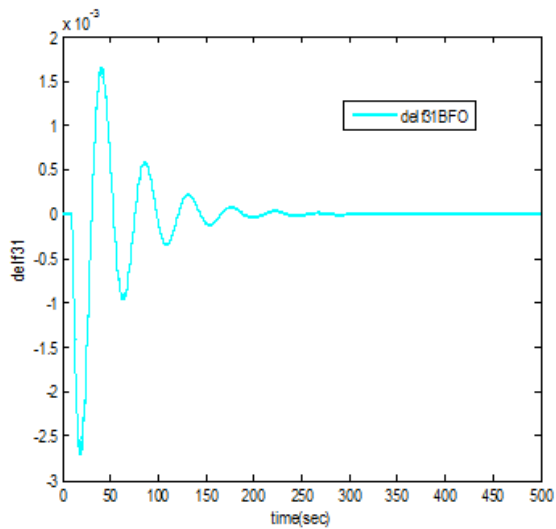

Fig 5. The frequency deviation in area 3 with 1% load change in area 1 with BFO technique

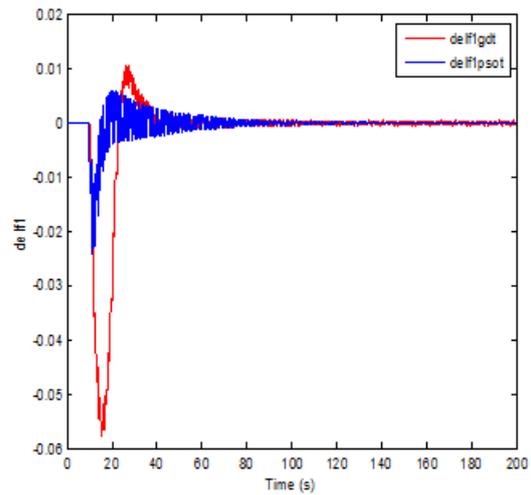

Fig 6. Frequency response for area-1 when 1% change in load of area-1

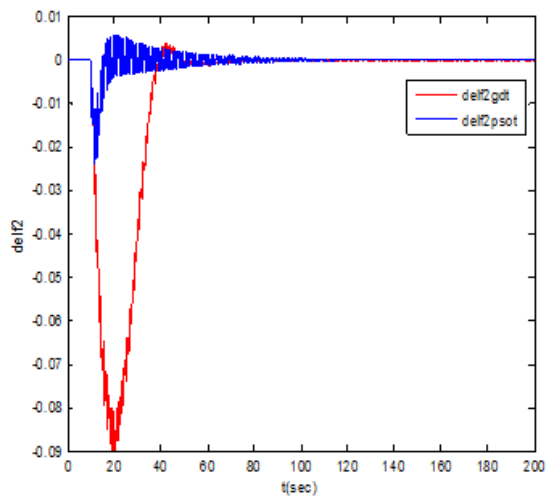

Fig 7. Frequency response for area-2 when 1% variation in load of area-1

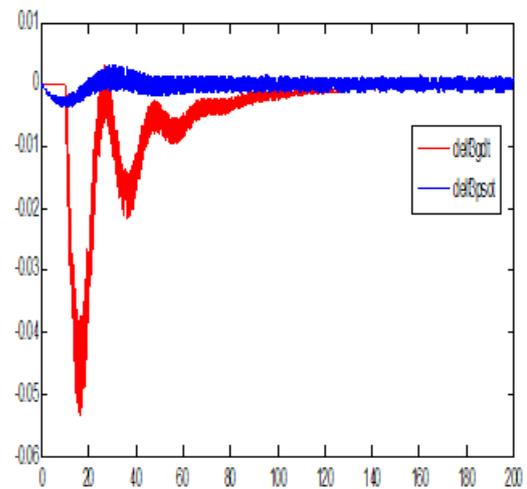

Fig 8. Frequency response for area-3 when 1% variation in load of area-1

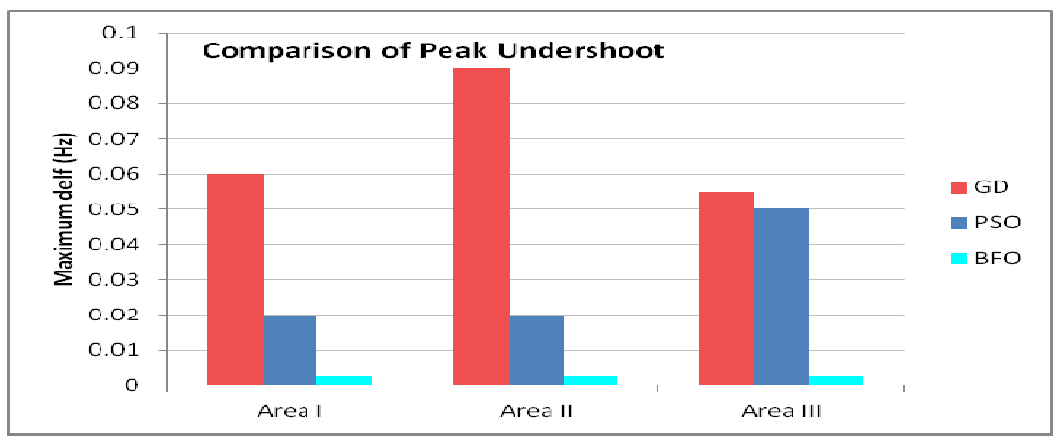

Fig. 9: Peak undershoot with GD, PSO and BFO methods with 1% load change in Area I

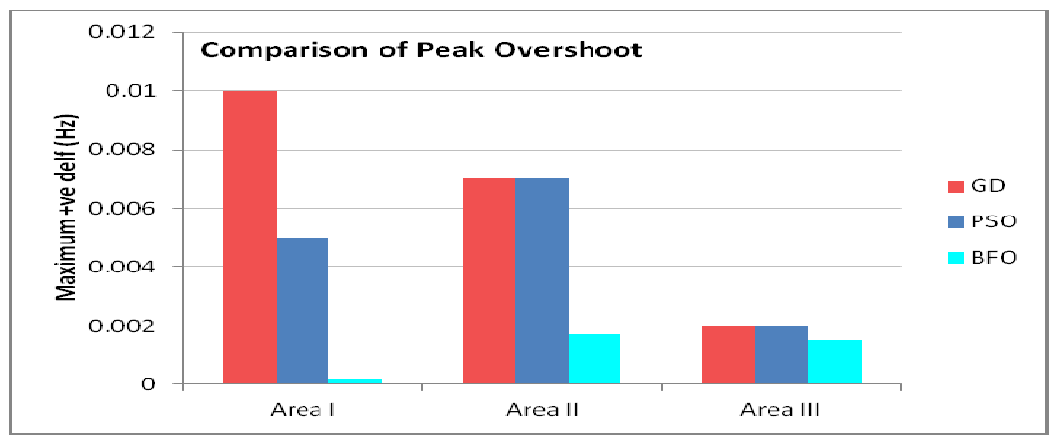

Fig.10: Peak Overshoot with GD, PSO and BFO methods with 1% load change in Area I

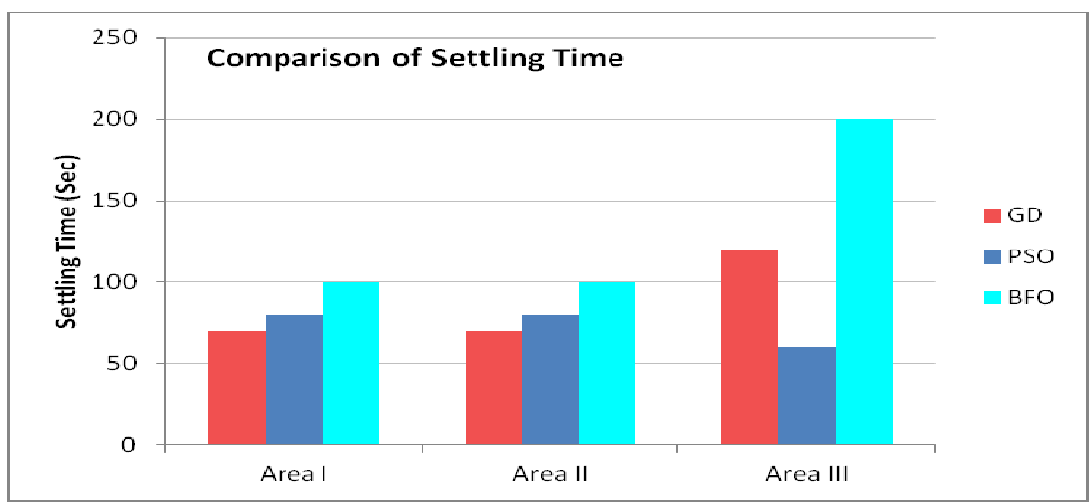

Fig 11:  Settling time with GD, PSO and BFO methods with 1% load change in Area I

It is observed that the peak overshoot and undershoot of the response decrease significantly with BFO followed by PSO and GD techniques as shown in Table III and IV . Although the settling time increases with BFO as given in Table V due to large number of the steps involved in BFO but this increase in settling time is marginal.

## V. CONCLUSION

The complex interconnected power system network with different generation systems has many parameters to be optimized which causes the difficulty in achieving the optimum frequency response and fast convergence. The controllers tuned with artificially intelligent techniques are effective for such type of systems. For the system investigated in this work total of twelve parameters have been optimized simultaneously and frequency response of the system is observed. The computationally intelligent techniques used in this work are BFO and PSO .One classical technique GD is also applied to observe the relative effectiveness of the computationally intelligent methods.

The best method among three methods which have been used in the present work is the BFO technique followed by PSO and GD methods. The effectiveness of these methods has been compared on the basis of settling time, peak overshoot and peak undershoot of the frequency response.

## APPENDIX

TABLE 1. Parameters of thermal and hydro systems

| | |
|---|---|
| f=60 Hz | Pr=2000MW |
| H=5 sec | Tr=10s, T12=0.544 |
| Ptie max=200 MW | Kp=120 Hz/puMW |
| Tt = 0.3sec | Tp=20 s |
| Tg= 0.08 s | Kr= 0.5 |
| D=0.00833 p.u.MW/Hz | R=2.4Hz/p.u.MW |

TABLE II. Parameters of Wind power plant

| | |
|---|---|
| Air density = 1.25 kg/m3 | $T_i$ = 3s |
| H=5 sec | $K_{pt}$ = 0.012 |
| Average wind velocity= 7 m/s | $T_{pt}$ = 10.55 |
| Radius of turbine blade=45 m Gear ratio =70 | Tp=20 s |

TABLE III. Peak Undershoot (Hz) with different methods

| Method | Area I | Area II | Area III |
|---|---|---|---|
| **GD** | 0.06 | 0.09 | 0.055 |
| **PSO** | 0.02 | 0.02 | 0.05 |
| **BFO** | 0.0027 | 0.0025 | 0.0025 |

TABLE IV. Peak Overshoot (Hz) with different methods

| Method | Area I | Area II | Area III |
|---|---|---|---|
| **GD** | 0.01 | 0.007 | 0.002 |
| **PSO** | 0.005 | 0.007 | 0.002 |
| **BFO** | 0.0002 | 0.0017 | 0.0015 |

TABLE V. Settling Time (sec) with different methods

| Method | Area I | Area II | Area III |
|---|---|---|---|
| **GD** | 70 | 70 | 120 |
| **PSO** | 80 | 80 | 60 |
| **BFO** | 100 | 100 | 200 |

## AUTHOR PROFILE

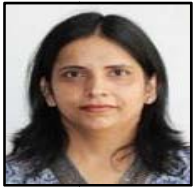
**Naresh Kumari** received her B.E. in Electrical Engineering with Honours in 1996 from C.R. State College of Engineering, Murthal, Haryana. She did her M. Tech. in Instrumentation and Control from Rajasthan University  in 2005 .She is currently pursuing her PhD under the supervision of Prof. A.N.Jha and Dr. Nitin Malik from The NorthCap University (NCU), Gurgaon, India. She is also working as Senior Assistant Professor in Department of Electrical, Electronics and Communication Engineering, The NCU, Gurgaon. She has 18 years of teaching and research experience. Her research interests include power system stability, renewable energy sources, intelligent control and wind energy conversion systems.

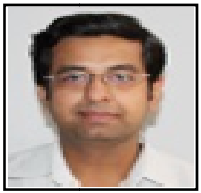
**Nitin Malik** received his B.Eng. (Electrical Engineering) from Chhotu Ram (C.R) State College of Engineering (now, Deenbandhu C.R. University of Science and Technology) in1999, M.Eng. (Industrial System & Drives) from MITS, Gwalior in 2001 and Ph.D (Power Systems) from Jamia Millia Islamia, New Delhi in 2014. He is Associate Professor of Electrical Engineering at The NorthCap University, Gurgaon, India. His research interest includes soft computing applications to Electric Power Distribution System security analysis,optimization and control. He has published 19 papers in International and National Journals andConferences and is supervising 4 PhD research scholars.

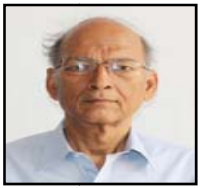
**A.N.Jha** has worked as Senior Professor in Department of Electrical, Electronics and Communication Engineering, The NCU, Gurgaon, India. He was Professor of Electrical Engineering in IIT, Delhi, India before joining The NCU, Gurgaon, India. He received his B.Sc. in Electrical Engineering from Bihar University, India in 1965 and M.E. (Electrical Engineering) in 1967 from University of Calcutta, India. He received his Ph.D. in Control System Engineering (Electrical) in 1977 from IIT Delhi, India. He has more than 40 years of teaching and research experience and has published/presented more than 120 papers in international and national journals/conferences. He has guided 70 M. Tech. and 7 Ph.D. students in their theses/projects.

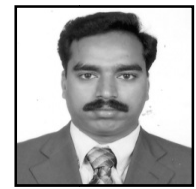
**G Mallesham** received the B.E degree in Electrical and Electronics Engineering from University College of Engineering, Osmania University, Hyderabad, and his Masters degree in Control Engineering and Instrumentation from Indian Institute of Technology, Delhi, India. He has done his  PhD  from Indian Institute of Technology, New Delhi, India and Post Doc from Yale University,USA. His interests are in soft computing, control, power systems and renewable energy.